\documentclass[%
 reprint,longbibliography,preprintnumbers,
nofootinbib,
 amsmath,amssymb,
 aps,
prl,
]{revtex4-2}
\pdfoutput=1
\usepackage[utf8]{inputenc}
\usepackage{flushend}
\usepackage{dcolumn}
\usepackage{bm}
\usepackage{balance}

\usepackage[normalem]{ulem}

\usepackage[colorlinks = true,
            linkcolor = blue,
            urlcolor  = blue,
            citecolor =green,
            anchorcolor = blue]{hyperref}
\usepackage{verbatim}
\usepackage{color,ulem}
\usepackage[english]{babel}

\usepackage[utf8]{inputenc}
\input Starburst.fd
\newcommand*\initfamily{\usefont{U}{Starburst}{xl}{n}}\initfamily

\newcommand{\beq}{\begin{eqnarray}}
\newcommand{\eeq}{\end{eqnarray}}
\usepackage{amsmath}
\usepackage{tikz}
\usetikzlibrary{decorations.pathmorphing}
\usetikzlibrary{shapes.misc}
\tikzset{cross/.style={cross out, draw=black, minimum size=8*(#1-\pgflinewidth), inner sep=0pt, outer sep=0pt},
cross/.default={1pt}}
\usetikzlibrary{patterns,math}
\begin{document}

\title{Explicit analytical solution for random close packing in $d=2$ and $d=3$}

\author{\textbf{Alessio Zaccone}$^{1,2}$}%
 \email{alessio.zaccone@unimi.it}
 
 \vspace{1cm}
 
\affiliation{$^{1}$Department of Physics ``A. Pontremoli'', University of Milan, via Celoria 16,
20133 Milan, Italy.}
\affiliation{$^{2}$Cavendish Laboratory, University of Cambridge, JJ Thomson
Avenue, CB30HE Cambridge, U.K.}

\begin{abstract}
We present an analytical derivation of the volume fractions for random close packing (RCP) in both $d=3$ and $d=2$, based on the same methodology. Using suitably modified nearest neigbhour statistics for hard spheres, we obtain $\phi_{\mathrm{RCP}}=0.65896$ in $d=3$ and $\phi_{\mathrm{RCP}}=0.88648$ in $d=2$. These values are well within the interval of values reported in the literature using different methods (experiments and numerical simulations) and protocols. This order-agnostic derivation suggests some considerations related to the nature of RCP: (i) RCP corresponds to the onset of mechanical rigidity where the finite shear modulus emerges, (ii) the onset of mechanical rigidity marks the maximally random jammmed state and dictates $\phi_{\mathrm{RCP}}$ via the coordination number $z$, (iii) disordered packings with $\phi>\phi_{\mathrm{RCP}}$ are possible at the expense of creating some order, and $z=12$ at the FCC limit acts as a boundary condition.

\end{abstract}

\maketitle

The problem of the closest packing of equal spheres was conjectured by Kepler for ordered assemblies of spheres. Gauss proved that the highest packing fraction that can be achieved by any packing of equal spheres is $\phi=\frac{\pi}{3\sqrt{2}} = 0.74048$, which corresponds to the face centered cubic (FCC) arrangement of spheres with coordination number $z=12$. The formal proof of Kepler's conjecture was provided by Hales in more recent times \cite{Hales1,Hales2,Hales3}.
An open central problem in contemporary physics and mathematics is the determination of the highest packing fraction occupied by spheres in disordered assemblies, i.e.  the so-called random close packing (RCP) problem. This problem is crucial for our understanding of amorphous materials~\cite{kob_book,stachurski}.
A visionary experiment by Bernal and Mason in 1960 \cite{Bernal} showed that random packings of spheres in $d=3$ have a volume fraction around $\phi_{\mathrm{RCP}}\approx 0.64$ and a coordination number $z=6$. This crucial observation was already understood by Bernal as a necessary consequence of \emph{mechanical stability}, since it was well known, since Maxwell, that a lattice with only central-force interactions is rigid only if $z \geq 6$, regardless of the lattice structure. 
Building on this intuition, Bernal speculated that tetrahedral arrangements of spheres must be dominant in RCP, since a given sphere needs to touch three nearest neighbours on a plane in order to be mechanically stable along the direction orthogonal to the plane. Bernal's idea prompted Finley and Gotoh \cite{Finley} to produce a heuristic analytical estimate of $\phi_{\mathrm{RCP}}$ in $d=3$ based on the local tetrahedral packing geometry. Many other approaches, in particular using computer simulations, are available ~\cite{Torquato_review}.

In $d=3$ the analytical approaches to the RCP problem are very few: besides the result of Finney and Gotoh we should mention
the analytical estimate obtained in Ref.\cite{Song2008} using the Edwards statistical mechanics of a restricted volume ensemble, the granocentric model of Ref. \cite{Jasna2009} based on coordination number and local available space, and the liquid-like approach based on metastability of Ref. \cite{Kamien}.

In $d=2$, a number of analytical estimates are available which are based on heuristic geometric considerations~\cite{Stillinger1964,Sutherland,Makse2014,Rafi}, while in high dimensional space theoretical descriptions provided by replica-symmetry breaking approaches become exact in the limit $d \rightarrow \infty$~\cite{Charbonneau}.

In general, an unambiguous determination of $\phi_{\mathrm{RCP}}$ is plagued by at least two aspects: (i) $\phi_{\mathrm{RCP}}$ depends largely on the protocol used to form the packing, and (ii) it is difficult to provide a clear-cut definition of ``randomness'' of the packing.
The latter point has been duly emphasized in \cite{Truskett}, where authors proposed that RCP is actually a maximally random jammed state corresponding to some minimum value of a structural order parameter. Another possible solution to this problem has been suggested in \cite{Kamien} with the idea that RCP is a singularity in a set of metastable
branches of the pressure, which echoes the high-dimensional findings of the replica method \cite{Charbonneau}.
Regarding the protocol dependence, this is manifested in the relatively broad range of $\phi_{\mathrm{RCP}}$ that have been reported in the literature, e.g. $\phi_{\mathrm{RCP}}=0.60 - 0.69$ in $d=3$ \cite{Torquato_review}, and $\phi_{\mathrm{RCP}}=0.81-0.89$ in $d=2$ \cite{Rafi}.

A simple theory that is able to predict $\phi_{\mathrm{RCP}}$ 
in both $d=2$ and $d=3$ has been missing due to 
the problem of analytically dealing with strong particle correlations (due to many-body excluded-volume interactions), which
preclude the development of a simple analytical theory.

Here we remedy to this situation and present a simple, analytical theory of RCP which, within the same framework, is able to predict sensible values for $\phi_{\mathrm{RCP}}$ in both $d=2$ and $d=3$. The approach and the derivation emphasize the role of mechanical stability and the emergence of rigidity in determining the RCP features.

We start with an operative definition of RCP based on mechanical stability.
The elasticity problem of random packings in $d=2$ and $d=3$ was solved exactly in Ref.\cite{Scossa}, and provides an accurate closed-form expression for the shear modulus:
\begin{equation}
G= \alpha(d)\, \rho\, \kappa\, \sigma^{2}\, (z-2d)
\label{scossa}
\end{equation}
where $\alpha(d=3) = \frac{1}{30}$ and $\alpha(d=2) = \frac{1}{18}$, $\rho=\frac{N}{V}$ is the particle density (obviously, $\rho=\frac{N}{S}$ in $d=2$), and $\kappa$ is the spring constant of the nearest-neighbour interaction, while $\sigma$ is the particle diameter. Equation \ref{scossa} was shown in Ref.\cite{Scossa} to be in excellent parameter-free agreement with simulations data of jammed packings from Ref. \cite{OHern}.

The negative contribution $\propto -2d$ arises from nonaffine motions of the particles under an applied shear strain. These motions arise in order to maintain the mechanical equilibrium on each particle, and represent a negative contribution to the free energy of deformation of the system. The dependence on space dimension $d$ is due to the fact that nonaffine relaxations involve all the degrees of freedom of the system, which are $dN$. The positive contribution is the affine Born-Huang contribution which is instead proportional to the total number of interparticle contacts, $\frac{zN}{2}$, hence the dependence on $z$.

Clearly, based on \eqref{scossa}, mechanical stability arises at $z=2d$ where the particle contacts become able to balance the energy cost of nonaffine relaxations. For $z<2d$ the system is not rigid, hence it is still able to undergo substantial rearrangements and to find denser configurations at larger $\phi$. At $z>2d$, instead, the system becomes jammed, and therefore $z=2d$  represents the point at which the disordered branch of the hard sphere state diagram must terminate. 

We therefore adopt Bernal's view of RCP and argue that $z=2d$  is the only well defined criterion to define RCP, whereas $\phi_{\mathrm{RCP}}$ follows from the $z=2d$ condition and is affected by the system- and protocol-specific ways by which $z=2d$ is reached.

In the following we therefore impose that $z=2d$ defines the RCP state in any dimension $d$, and we derive $\phi_{\mathrm{RCP}}$ from this condition for $d=3$ first and subsequently double check that the same procedure yields a sensible estimate of $\phi_{\mathrm{RCP}}$ also in $d=2$.

To deal with the strong statistical correlations among particles in the dense hard sphere system, we employ suitably modified liquid state theory for the radial distribution function (rdf). It is known that liquid theories of the rdf are unable to predict the divergence of pressure at RCP, and also cannot predict the formation of permanent nearest-neighbour contacts at RCP. However, they still provide a useful analytical starting point to account for the statistical increase of crowding around a test particle, as $\phi$ increases \cite{Torquato_1995}. 

We start from the standard definition of coordination number $z$ based on the rdf, which in $d=3$ reads as~\cite{Hansen}:
\begin{equation}
\mathrm{d} z = 4\pi \, \rho g(r) \, r^{2} \, \mathrm{d}r
\end{equation}
where $\mathrm{d} z$ represents the average number of particles lying in the range $r + \mathrm{d}r$. 

We now introduce the quantity $\sigma^{+} \equiv \sigma + \epsilon$ where $\epsilon$ is an arbitrarily small number, $\epsilon \rightarrow 0^{+}$.
Hence the average number of particles in contact (just touching) with a test particle is given (in $d=3$) by
\begin{equation}
z = 4\pi \rho \int_{0}^{\sigma^{+}}g(r) r^{2} \mathrm{d}r.
\end{equation}

As is standard, $g(r)$ is defined as the probability of finding the centre
of a particle at a distance $r$, within $\mathrm{d}r$, from the test particle at the origin of the reference frame \cite{Hansen}. 
Focusing on the contact or near contact region, we define a suitably normalized probability density function (pdf) for the contact region $g_{c}(r)$ in a generic $d$-dimensional space:
\begin{equation}
\frac{\rho}{z}\int_{0}^{\sigma^{+}}g_{\mathrm{c}}(r)\mu(r)\mathrm{d}r=1
\label{norm}
\end{equation}
where $\mu(r)$ is the appropriate metric factor for the $d$-dimensional space, e.g. $\mu(r)=4\pi \, r^{2}$ in $d=3$, $\mu(r)=2\pi \, r$ in $d=2$ and so on.

In probability theory, besides fully continuous and fully discrete probability distribution functions, one can also define partially continuous distributions, also known as mixed distributions or mixed random variables \cite{Shynk}. As an example of a fully discrete distribution, the pdf $f_{\mathrm{d}}(x)$ of a distribution consisting of a set of points $x_{i}=\{x_1,...,x_{n}\}$, with corresponding probabilities $p_{i}=\{p_1,...,p_{n}\}$ can be written as $f_{\mathrm{d}}(x) = \sum_{i}p_{i}\delta(x-x_{i})$ \cite{Khuri}. A partially continuous distribution can be written as \cite{Shynk}: $f_{\mathrm{pc}}(x)=c(x) + \sum_{i}p_{i}\delta(x-x_{i})$ where $c(x)$ is the continuous part and the second term is the discrete part. The latter implies that the distribution returns exactly the value  $x_{i}$ with probability $p_{i}$.
Upon normalizing to 1 over the relevant domain, $\int_{0}^{\infty}f_{\mathrm{pc}}\,\mathrm{d}x=1$, this is indeed a valid pdf \cite{Hossein}.\\

Following the above considerations, we can treat the total $g(r)$ as a partially continuous pdf. Hence we split it into a discrete part which describes the probability of having nearest neighbours in direct contact with the test particle, that we call $g_{\mathrm{c}}(r)$, and a continuous part which describes the probability of finding particles in the region of space beyond contact (bc), i.e. $r > \sigma^{+}$, that we call $g_{\mathrm{bc}}(r)$,
\begin{equation}
g(r) = g_{\mathrm{c}}(r) + g_{\mathrm{bc}}(r).
\label{mixed}
\end{equation}
Here, $g_{\mathrm{c}}(r)$ is a discrete probability distribution defined (consistent with the above generic examples \cite{Khuri,Hossein}) as:
\begin{equation}
g_{\mathrm{c}}(r) = g_{0} \, g(\sigma) \, \delta(r-\sigma).
\label{key_result}
\end{equation}
where $g(\sigma)$ is the contact value of the $g(r)$ \cite{Hansen,Torquato_book}, i.e. the probability of finding particles at exactly $r=\sigma$, and $g_{0}$ is a normalization factor to be determined later from  \eqref{norm}. In turn, the total $g(r)$ is therefore a generalized pdf \cite{Hossein},which obeys the usual normalization condition $\frac{1}{N}\int_{0}^{\infty} 4\pi \, \rho g(r) \, r^{2} \, \mathrm{d}r = 1$.

The statistical theory of hard-sphere liquids provides a way to compute $g(\sigma)$ analytically up to the (unphysical) packing fraction $\phi=1$ \cite{Torquato_1995,Torquato_book,Stratt}, while remaining agnostic about the possible onset of ordering. 
Exact analytical solutions for $g(\sigma)$ of hard spheres in all odd space dimensions are available based on Percus-Yevick (PY) theory \cite{Leutheusser}. In $d=3$ the PY result is \cite{Leutheusser,Stratt}
\begin{equation}
g(\sigma) = \frac{1 + \frac{\phi}{2}}{(1-\phi)^{2}}
\label{PY}
\end{equation}
and alternatively one can use the very accurate Carnahan-Starling (CS) expression \cite{Stratt}:
\begin{equation}
g(\sigma) = \frac{1 - \gamma(d)\phi}{(1-\phi)^{d}}.
\label{CS}
\end{equation}
with $\gamma(d=3)=0.5$ and $\gamma(d=2) = 0.43599$.

As the last step, we now only need a condition to determine the normalization factor $g_{0}$ in the definition of $g_{\mathrm{c}}$. First of all, we notice that based on dimensional analysis, $g_{0} \propto \sigma$, to ensure dimensional consistency of Eq.\eqref{norm}.
The only possible condition that we can choose to determine the numerical prefactor in $g_{0} \propto \sigma$ is based on the closest packing (CP) value which spherical objects can never exceed. As is well known, the closest packing of spheres in $d=3$ occurs when $z=12$, and $\phi_{\mathrm{CP}}=\frac{\pi}{3\sqrt{2}}=0.74048$ \cite{Hales2,kob_book}. Of course at this point the system has perfect (FCC) ordering. This limit can be used as an effective`` boundary condition'' in our problem to determine the unknown prefactor. This choice is consistent with the well known fact that disordered packings with partial order can be formed in the range $\phi_{\mathrm{RCP}} \phi < \phi_{CP}$ \cite{Truskett}.

Using therefore $z=12$ and $\phi_{\mathrm{CP}}  = 0.74048$ in \eqref{norm}, and recalling that $\frac{4}{3}\left(\frac{\sigma}{2}\right)^{3}\rho \equiv \phi$, in $d=3$, we just need to solve:
\begin{equation}
24 \, \phi_{\mathrm{CP}} \, \frac{1}{\sigma}\, g_{0}\, g(\sigma) = 12
\end{equation}
where $g(\sigma)$ is evaluated at $\phi_{\mathrm{CP}}=\frac{\pi}{3\sqrt{2}}=0.74048$. Using the PY expression \eqref{PY}, we can solve analytically for $g_0$ and find
\begin{equation}
g_{0} = \frac{(\sqrt{2} \pi - 6)^2}{\sqrt{2} \pi (\sqrt{2} \pi + 12)} \, \sigma \approx 0.0331894\, \sigma.
\end{equation}

We can now insert this result in \eqref{key_result}, and replace the latter in \eqref{norm}.
We now impose the RCP condition $z=2d=6$ valid in $d=3$ from \eqref{scossa}, and solve analytically the following equation
\begin{equation}
24 \, \frac{(\sqrt{2} \pi - 6)^2}{\sqrt{2} \pi (\sqrt{2} \pi + 12)} \, \phi_{\mathrm{RCP}} \, \frac{1 + \frac{\phi_{\mathrm{RCP}}}{2}}{(1-\phi_{\mathrm{RCP}})^{2}} = 6
\end{equation}
from which we obtain the explicit analytical solution for the random close packing fraction in $d=3$:
\begin{equation}
\begin{split}
\phi_{\mathrm{RCP}}^{(3\mathrm{D})} =&\frac{2\sqrt{648+\pi[\,\pi(54-24\sqrt2\pi+5\pi^2)-108\sqrt2\,]}}{36\sqrt2+\pi(\sqrt2\pi - 36)}+\\ &+\frac{2(36\sqrt2-48\pi)}{36\sqrt2+\pi(\sqrt2\pi-36)}-3\\ &=0.658963
\end{split}
\end{equation}
which is well within the range $0.61-0.69$ for RCP observed with different experiments and simulations \cite{Torquato_review}, and somewhat larger but not too far from the most quoted value $\phi \approx 0.64$.\\

The same procedure can be done using the CS expression instead of the PY one, which yields $\phi_{\mathrm{RCP}}=0.677376$, i.e. a higher value. This can be understood as a different ``protocol'' for implementing statistical strong correlations among particles. One should note that while $z=2d$ always applies at RCP, the corresponding $\phi_{\mathrm{RCP}}$ is not univocally determined and depends on the realization of disorder and the ``crowding'' protocol.\\

As a consistency check, we now turn to the RCP in $d=2$. The metric factor in \eqref{norm} is now $\mu(r) = 2 \pi r$, and we need to resort to the CS expression valid in $d=2$ for $g(\sigma)$, provided by \eqref{CS} with $d=2$.
The condition that we need to apply to determine the numerical prefactor of $g_0$ is analogous, \emph{mutatis mutandis}, to the one used in $d=3$. That is, we impose the maximum close packing condition in $d=2$, which is $z=6$ at $\phi_{\mathrm{CP}}^{(2\mathrm{D})}=\frac{\pi}{2 \sqrt{3}}=0.90690$.
Proceeding in the same manner as before, this time we obtain:
\begin{equation}
8 \, \phi_{\mathrm{CP}} \, \frac{1}{\sigma}\, g_{0}\, g(\sigma) = 6
\end{equation}
and again $g_0 \propto \sigma$ for dimensional reasons. The above equation can be solved analytically for the prefactor of $g_{0}$, by imposing $\phi_{\mathrm{CP}}^{(2\mathrm{D})}=\frac{\pi}{2 \sqrt{3}}=0.90690$, which yields:
\begin{equation}
g_{0} = 0.011856 \, \sigma.
\end{equation}

As done before, we can now insert this result in \eqref{key_result}, and replace the latter in \eqref{norm}. We now impose the RCP condition $z=2d=4$ from \eqref{scossa}, and solve analytically the following equation
\begin{equation}
8 \cdot 0.011856 \cdot \phi_{\mathrm{RCP}}\frac{1 - 0.43599\phi_{\mathrm{RCP}}}{(1-\phi_{\mathrm{RCP}})^{2}} = 4
\end{equation}
from which we obtain the explicit solution for the random close packing in $d=2$:
\begin{equation}
\phi_{\mathrm{RCP}}^{(2\mathrm{D})} = 0.88644.
\end{equation}
This estimate is again within the widely reported interval $0.81 - 0.89$ \cite{Rafi}, although closer to the upper end. 
Values of $\phi_{\mathrm{RCP}} \approx 0.89$ have been reported in numerical simulations \cite{Makse_2010} and analytical estimates based on heuristic local packing geometry considerations \cite{Stillinger1964}.

In future work, this approach can be further extended in several directions. For example, it can be extended to non-spherical particle packings, e.g. packings of ellipsoids. This is possible because (anisotropic) expressions for the $g(\mathbf{r})$ based on PY theory are available also for ellipsoids \cite{Latz}, and have been used in the past to study glass transition of dumbbells within mode-coupling theory \cite{Goetze_ellipsoid} and numerical simulations \cite{Sciortino}. The extended PY theory of Ref. \cite{Latz} can thus be used as input within the above framework to make explicit predictions for the $\phi_{\mathrm{RCP}}$ of ellipsoids. Similarly, the present approach can also be extended to higher spatial dimensions, by using PY theory or CS expressions valid for higher dimensions \cite{Stratt}, together with suitably modified expressions for the hypersphere packing fraction and volume integration metrics \cite{Conway}. This will lead to explicit formulae for $\phi_{\mathrm{RCP}}$ as a function of $d$  that can be compared with existing computationally more elaborate approaches \cite{Charbonneau}.

Further future extensions, in perspective, may lead to application of the present framework to systems with inherent contact network, like
freely jointed chains of hard spheres \cite{Laso} or even linear polymer chains of spheres where jamming (RCP) is closely related to glass transition \cite{Hoy}.

In summary, we presented the first simple and closed-form analytical solution for the random packing problem in both $d=2$ and $d=3$. While previous approaches rely on elaborate theoretical frameworks and often involve numerical steps to arrive at the final solution, or are based on heuristic local geometry considerations, the solution presented here relies exclusively on the statistical mechanics of hard spheres to account for the strong particle correlations and the increase of crowding upon increasing the packing fraction. 
The derivation is therefore ``order-agnostic'' in the sense that it does not specify the structural ordering of the particles but merely their excluded volume correlations. 
As suggested by the solution procedure, the only well defined notion of RCP is given in terms of the coordination number, which is $z=2d$ at RCP in $d$ dimensions. As shown above, this is the only well defined starting point to compute the packing fraction at RCP, which instead is not univocally defined and depends on the actual protocol that one uses to implement the spatial correlations between the particles. This is exemplified by the slightly different values of $\phi_{\mathrm{RCP}}$ using different implementations of hard sphere theory.
The new method introduced above can be easily extended in future work to dimensions $d > 3$ and to non-spherical packings.

\subsection*{Acknowledgments} 
We thank Matteo Baggioli and Mary Thomson for useful suggestions.
A.Z. acknowledges financial support from US Army Research Office, contract nr. W911NF-19-2-0055. 
\bibliographystyle{apsrev4-2}
\bibliography{refs}

\end{document}